\documentclass[referee]{aa}
\usepackage{graphics}
\begin{document}
\thesaurus{02(12.12.1)}
\title{The fractal octahedron network of the large scale structure}

\author{E. Battaner}
\institute{Dpto. Fisica Teorica y del Cosmos. 
Universidad de Granada. Spain}

\date{}

\maketitle

\begin{abstract}
In a previous article, we have proposed that the large scale structure
network generated by large scale magnetic fields could consist of a
network of octahedra only contacting at their vertexes. Assuming such a
network could arise at different scales producing a fractal geometry,
we study here its properties, and in particular how a sub-octahedron network
can be inserted within an octahedron of the large network. We deduce that
the scale of the fractal structure would range from $\approx$100 Mpc,
i.e. the scale of the deepest surveys, down to about 10 Mpc, as other smaller
scale magnetic fields were probably destroyed in the radiation
dominated Universe.

\end{abstract}

A fractal nature for the  large-scale
structure of the Universe has often been suggested
(early references in Mandelbrot, 1984;
Coleman \& Pietronero, 1992; Guzzo, 1997; Sylos Labini et al. 1998a,b;
Coles, 1998; de Vega et al., 1998 and others). On the other
hand, it has also been suggested (Battaner et al.,
1997a,b,  and Florido
\& Battaner, 1997) that a network built up of octahedra only
contacting at their
vertexes (like an egg carton) is the simplest network comprising
filaments produced by early large-scale magnetic fields. The purpose
of this paper is to show how both suggestions may be compatible, i.e. how to
produce a fractal based on filament octahedron elements, in other
words,  how to
insert the sub-octahedron network inside an octahedron belonging to
the larger scale network.

The size of the large octahedron can be 2, 3, 4, $\dots$ times the
size of the octahedra inside it. But let us exclude from this series
even ratios 2, 4, 6 $\dots$ because it is impossible to insert
octahedra contacting at their vertexes in such a way as to have a
sub-octahedron at each corner of the octahedron. The ratio of the
sizes of octahedron and suboctahedron is therefore assumed to be 3, 5,
7 $\dots$. The simplest corresponds to a ratio 3, which is depicted in
fig. 1. In this case, 7 sub-octahedra are contained in the octahedron.

\begin{figure}
 \resizebox{7cm}{!}{\includegraphics{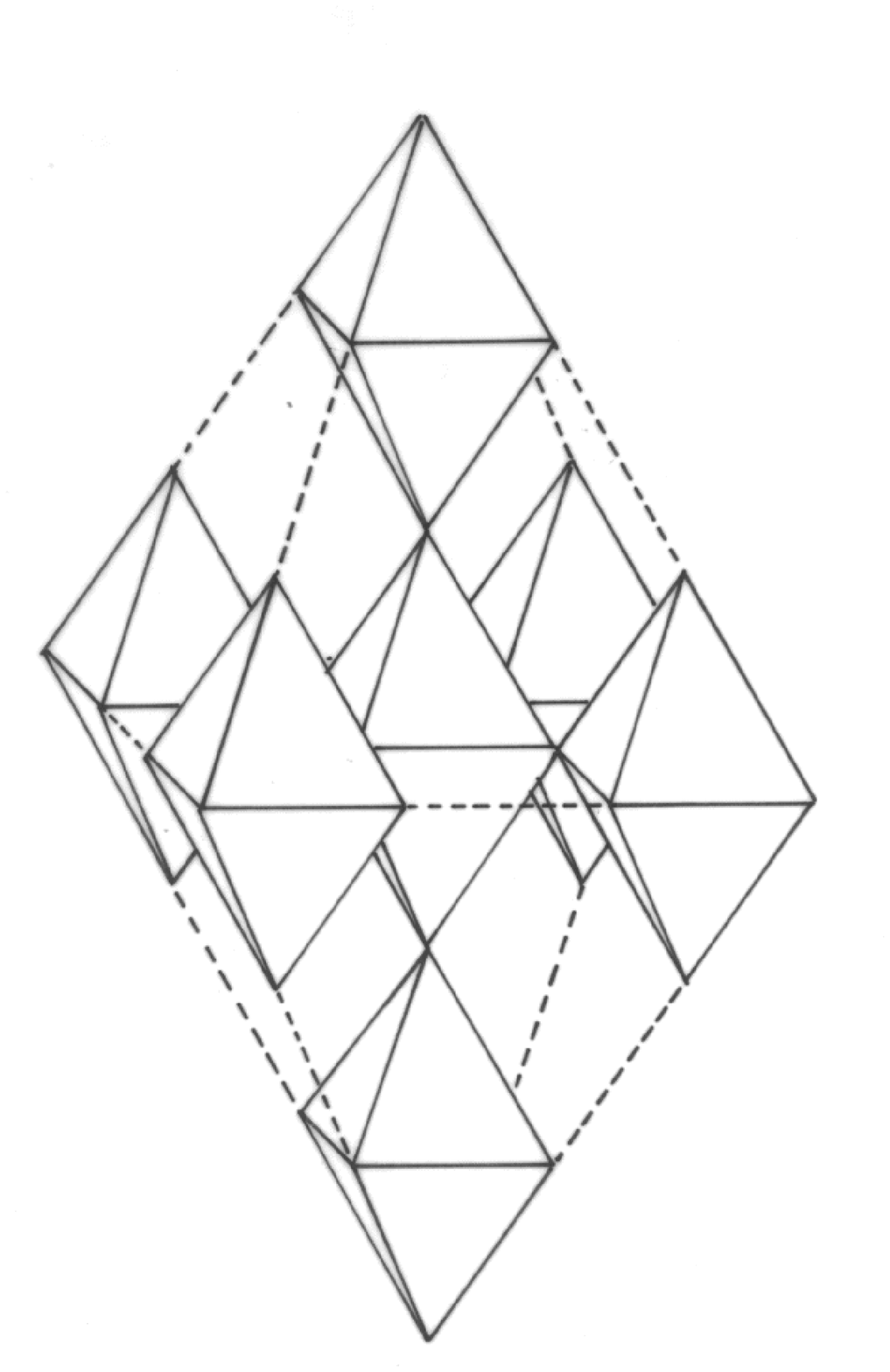}}
 \caption{7 small octahedra inside a large octahedron when the size
 ratio is 3}
 \label{}
\end{figure}

For a ratio of 5, there are 25 sub-octahedra. For a ratio of 7, there
are 63, and so on. Table 1 gives the series of
sub-octahedra that are contained in the large octahedron, N.

\begin{table}
\caption[]{}
\begin{flushleft}
\begin{tabular}{ccc}
\hline\noalign{\smallskip}
Order number & Ratio of sizes & Number of sub-octahedra\\
n            & 2n-1           &  N                     \\ 
\noalign{\smallskip}
\hline\noalign{\smallskip}
1 & 1 & 1 \\
2 & 3 & 7 \\
3 & 5 & 25 \\
4 & 7 & 63 \\
5 & 9 & 129 \\
6 & 11 & 231 \\
7 & 13 & 377 \\
\noalign{\smallskip}
\hline
\end{tabular}
\end{flushleft}
\end{table}

This series can be represented by the equation 

\begin{equation}
N(n)={4 \over 3} n^3 + 2 n^2 +{8 \over 3} n -1
\end{equation}

The fractal dimension would be 1.77, 2, 2.13, 2.21, $\dots$ ($\rightarrow$3
for n$\rightarrow \infty$) not considering mass distributions, but
rather lengths of filaments (smaller filaments contain smaller mass
objects, Lindner et al., 1996). The values of the fractal dimension
are similar to those obtained by other authors (see Einasto et
al. 1997 and references therein).

For n=3, we obtain a fractal dimension of 2, which is precisely the
value suggested by observations (Pietronero et al., 1996; Sylos Labini
et al., 1998) but values given in the above list are all compatible
with our network.

Lindner et al (1996) considered the range over which filamentary
structures exist, from the observational point of view. Following the
magnetic interpretation assumed here, the lower limit corresponds to
the smaller scale below which primordial magnetic fields were
destroyed in the radiation-dominated era by different mechanisms,
mainly Silk damping (Silk, 1968; Jedamzik, Katalinic and Olinto, 1996)
and magnetic diffusion (Lesch and Birk, 1998). Super horizon scale
magnetic fields during this epoch, today of the order of 10 Mpc, were
unaffected by these processes and this horizon scale should be
identified 
with the
lower limit, below which the fractal geometry is lost. The upper
limit should be greater than the typical distance of the deepest
surveys, about 100 Mpc. Therefore the distance range is relatively
short, 10-100 Mpc. In the best case with a size ratio equal to 3,
there would only be octahedra, sub-octahedra and sub-sub-octahedra,
nothing more. Even the lowest size octahedra would probably be
unrecognizable. We also conclude that $n \le 2$.
This situation is similar to that depicted by Lindner et
al (1996).

\begin{acknowledgements}
This paper has been supported by the spanish ``Ministerio de Educacion
y Cultura'' (PB96-1428) and the ``Plan Andaluz de Investigacion''
(FQM-0108).
\end{acknowledgements}

\end{document}